\documentclass[12pt]{article}

\usepackage{cite}
\usepackage{epsfig}
\usepackage{amsmath}
\usepackage{array}
\usepackage{multirow}
\def\mathswitch#1{\relax\ifmmode#1\else$#1$\fi}
\def\mathswitchr#1{\relax\ifmmode{\mathrm{#1}}\else$\mathrm{#1}$\fi}
\newcommand{\PW}{\mathswitchr W}
\newcommand{\PZ}{\mathswitchr Z}
\newcommand{\PH}{\mathswitchr H}

\newcommand{\Pb}{\mathswitchr b}
\newcommand{\Pt}{\mathswitchr t}

\newcommand{\MW}{\mathswitch {M_\PW}}
\newcommand{\MZ}{\mathswitch {M_\PZ}}

\newcommand{\MH}{\mathswitch {M_\PH}}

\newcommand{\mb}{\mathswitch {m_\Pb}}
\newcommand{\mt}{\mathswitch {m_\Pt}}

\newcommand{\as}{\alpha_{\mathrm s}}

\newcommand{\gev}{\,\, \mathrm{GeV}}

\newcommand{\mycaption}[1]{\caption{\sl #1}}

\makeatletter
\def\section{\@startsection {section}{1}{\z@}{-3.5ex plus -1ex minus 
 -.2ex}{2.3ex plus .2ex}{\large\bf\boldmath}}
\def\subsection{\@startsection{subsection}{2}{\z@}{-3.25ex plus -1ex
 minus -.2ex}{1.5ex plus .2ex}{\normalsize\bf\boldmath}}
\def\subsubsection{\@startsection{subsubsection}{3}{\z@}{-3.25ex plus
 -1ex minus -.2ex}{1.5ex plus .2ex}{\normalsize\it}}

\graphicspath{{.}{plots/}}

\begin{document}
\thispagestyle{empty}

\def\thefootnote{\fnsymbol{footnote}}

\begin{flushright}
\end{flushright}

\vspace{1cm}

\begin{center}

{\Large {\bf On the evaluation of two-loop electroweak box diagrams for $e^+e^- \to HZ$ production}}
\\[3.5em]
{\large
Qian~Song and Ayres~Freitas
}

\vspace*{1cm}

{\sl
Pittsburgh Particle-physics Astro-physics \& Cosmology Center
(PITT-PACC),\\ Department of Physics \& Astronomy, University of Pittsburgh,
Pittsburgh, PA 15260, USA
}

\end{center}

\vspace*{2.5cm}

\begin{abstract} 
Precision studies of the Higgs boson at future $e^+e^-$ colliders can help to
shed light on fundamental questions related to electroweak symmetry breaking,
baryogenesis, the hierarchy problem, and dark matter.
The main production process, $e^+e^- \to HZ$, will need to be controlled with
sub-percent precision, which requires the inclusion of next-to-next-to-leading
order (NNLO) electroweak corrections. The most challenging class of diagrams are
planar and non-planar double-box topologies with multiple massive propagators in
the loops. This article proposes a technique for computing these diagrams
numerically, by transforming one of the sub-loops through the use of Feynman
parameters and a dispersion relation, while standard one-loop formulae can be
used for the other sub-loop. This approach can be extended to deal with tensor
integrals. The resulting numerical integrals can be evaluated in minutes on a
single CPU core, to achieve about 0.1\% relative precision.
\end{abstract}

\setcounter{page}{0}
\setcounter{footnote}{0}

\newpage

%%%%%%%%%%%%%%%%%%%%%%%%%%%%%%%%%%%%%%%%%%%%%%%%%%%%%%%%%%%%%%

\section{Introduction}

Since the discovery of the Higgs boson in 2012 \cite{higgs1}, experiments at the
Large Hadron Collider (LHC) have measured many properties of this new particle,
in overall good agreement with the predictions of the Standard Model (SM)
\cite{rpp}. However, in most models beyond the SM, one would expect deviations
of the Higgs boson couplings at the per-cent level \cite{Englert:2014uua}, which is
beyond the achievable precision at the LHC.

For this reason, several proposals have been made for so-called $e^+e^-$ Higgs
factories, operating at center-of-mass energies of 240--250~GeV: the International
Linear Collider (ILC) \cite{ilc}, the Future Circular Collider (FCC-ee)
\cite{fccee}, and the Circular Electron-Positron Collider (CEPC) \cite{cepc}.
These machines would be able to study the Higgs boson through the process
$e^+e^- \to HZ$ in a clean environment and produce per-cent level precision
measurements of the dominant Higgs couplings. The $HZZ$ coupling can be
extracted from the $HZ$ production cross-section itself, $\sigma_{HZ}$, which is
anticipated to be measurable with a precision of about 1.2\% at ILC, 0.4\% at
FCC-ee, and 0.5\% at CEPC.

The interpretation of $\sigma_{HZ}$ in terms of the $HZZ$ coupling requires
precise theoretical predictions for the process $e^+e^- \to HZ$, including
radiative corrections. The next-to-leading order (NLO) corrections within the SM
have been known since a long time for unpolarized beams \cite{nlo}, and more
recently for polarized beams \cite{Bondarenko:2018sgg}. Mixed electroweak--QCD
${\cal O}(\alpha\as)$ corrections have been computed in Ref.~\cite{ewqcd},
which required the evaluation of two-loop self-energy and vertex diagrams. Given
the relatively large decay width of the Z boson, the predictions can be further
refined by including the Z-boson decay at the same order, \emph{i.e.} by
computing corrections to the process $e^+e^- \to Hf\bar{f}$. The NLO electroweak
contributions to this process for the final states $f=\nu_e$ and $f=e$ have been
studied in Ref.~\cite{nlo2}, and the ${\cal O}(\alpha)$ and ${\cal O}(\alpha\as)$ corrections for
$f=\mu$ have also become available recently \cite{ewqcd2}.

The numerical impact of the one-loop corrections is at the level of 5--10\%,
with a dominant contribution stemming of initial-state radiation (ISR) of soft
and collinear photons. These ISR effects are enhanced by logarithmic terms of the form
$\log(s/m_e^2)$. In the soft and collinear limit, these logarithmic terms are
process-independent, and higher-order ISR contributions can be included through
the structure function method \cite{strfunc}. The impact of ISR on $e^+e^- \to
HZ$ has been recently studied \cite{isrhv,isrhv2}. It was found that, when
including third-order corrections in the structure function \cite{strfunc3}, the
uncertainty from missing higher ISR orders is at the level of $10^{-4}$ and thus
negligible \cite{isrhv2}.

The ${\cal O}(\alpha\as)$ contributions modify the $HZ$ cross-section by about
1.5\% when parametrizing the elecroweak couplings in terms of $\alpha$, and
about 0.4\% using the Fermi constant $G_\mu$ instead. These corrections are
sizeable and must be taken into account for analyses at future $e^+e^-$ Higgs
factories. The largest unknown higher-order contribution stems from electroweak
two-loop effects, which are expected to have an impact at the level of ${\cal
O}(1\%)$ \cite{therr}, and thus also have to be included. Given that the
width-to-mass ratio of the Z boson, $\Gamma_Z/m_Z \sim 2.7\%$, is comparable to
one order in electroweak perturbation theory, it is sufficient to compute these
NNLO electroweak corrections for the process $e^+e^- \to HZ$ with an on-shell
Z boson, whereas the full process $e^+e^- \to Hf\bar{f}$ should be
treated at the NLO level. These two contributions can be consistently
combined by performing an expansion about the pole of the Z boson \cite{zpole}.

Among the NNLO electroweak corrections, diagrams with closed fermion loops are 
typically dominant, due to the large top-quark Yukawa coupling and the large
multiplicity of fermions in the SM\footnote{This expectation is corroborated by
known examples of NNLO electroweak calculations, see \emph{e.g.}
Ref.~\cite{Dubovyk:2019szj}.}. Within this class of diagrams, the most
challenging piece are planar and non-planar two-loop box graphs with top quarks
inside one sub-loop\footnote{Diagrams without top quarks are negligible due to
the small fermion Yukawa couplings.}. Even when neglecting
all fermion masses besides the top quark, these contributions depend on up to
four independent mass scales ($m_H,\,m_Z,\,m_W,\,m_t$), as well as two
additional momentum scales (which can be represented by the Mandelstam variables
$s$ and $t$). Therefore it is difficult to find analytical solutions,
since the expressions will be impractically large and may require the
development of new special functions. On the other hand, generic numerical methods (such
as numerical integration over Feynman parameters \cite{Yuasa:2011ff}) are
highly computationally intensive. 

In this paper, a more efficient numerical method for the evaluation of two-loop
box integrals is proposed. It is based on a combination of a dispersion relation
and Feynman parameters for one of the two sub-loops \cite{Awramik:2006uz}. The
method of Ref.~\cite{Awramik:2006uz} is extended to enable the direct evaluation
of tensor integrals (rather than attempting to reduce them to a set of master
integrals)\footnote{See Ref.~\cite{Aleksejevs:2018tfr} for a similar technique
for tensor integrals, which however differs in several technical
details.}.  This approach leads to three-dimensional numerical integrals for the
two-loop boxes, which can be evaluated with about four-digit precision within
minutes on a single CPU core.

In the following section, the derivation of the numerical integral
representations for the planar and non-planar two-loop box diagrams is discussed
in detail. Section~\ref{res} describes the application of this method to the
evaluation of two-loop box diagrams contributing to the process $e^+e^- \to HZ$,
including several important aspects of the numerical implementation, as well as
numerical results for these diagrams. The main findings of this paper are
summarized in section~\ref{sum}.

%%%%%%%%%%%%%%%%%%%%%%%%%%%%%%%%%%%%%%%%%%%%%%%%%%%%%%%%%%%%%%

\section{Evaluation of two-loop box diagrams with a top loop}

\subsection{Planar diagrams}
\label{plan}

%-------------------------------------------------------------
\begin{figure}[tb]
\centering
\begin{tabular}{ccc}
\epsfig{figure=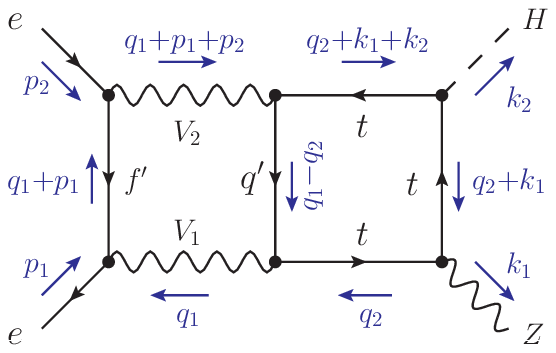, bb=150 540 450 710, clip=1, height=1.5in} 
 & \hspace{.5in} &
\epsfig{figure=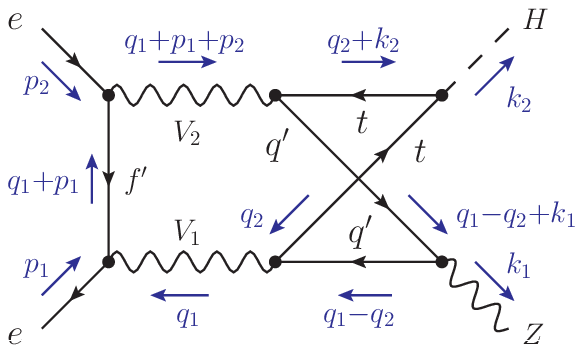, bb=150 540 450 710, clip=1, height=1.5in} \\
{\LARGE $\downarrow$} && {\LARGE $\downarrow$} \\[1em]
\epsfig{figure=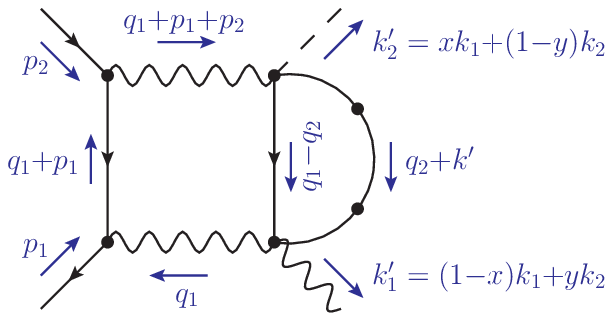, bb=180 540 480 710, clip=1, height=1.5in}
 & \hspace{.5in} &
\epsfig{figure=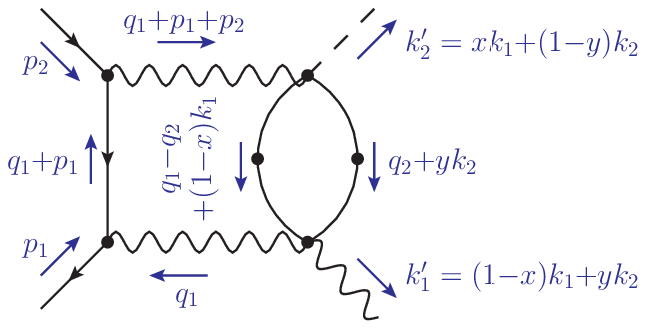, bb=180 540 480 710, clip=1, height=1.5in}
\end{tabular}
\mycaption{Planar (left) and non-planar (right) two-loop box diagrams with top
quarks in the loop. The bottom row visually illustrates the effect of
introducing Feynman parameters for the top loop. If $V_{1,2}=\gamma,Z$ then
$f'=e$, $q'=t$, whereas $f'=\nu_e$ and $q'=b$ for $V_{1,2}=W$.}
\label{fig:diags}
\end{figure}
%-------------------------------------------------------------

To illustrate the method, let us initially consider a basic scalar planar
integral, which contains the propagators of the diagram in
Fig.~\ref{fig:diags} (top-left)
but simply 1 in the numerator:
\begin{align}
I_{\rm plan} = \int &d^Dq_1 \, d^Dq_2 \; \frac{1}{[q_1^2-m_{V_1}^2]
 [(q_1+p_1)^2-m_{f'}^2][(q_1+p_1+p_2)^2-m_{V_2}^2]} \notag \\
&\times \frac{1}{[(q_1-q_2)^2-m_{q'}^2][q_2^2-\mt^2][(q_2+k_1)^2-\mt^2]
 [(q_2+k_1+k_2)^2-\mt^2]}\,.
\end{align}
The extension to non-trivial tensor structures in the numerator will be
discussed below.

The following approach is based on the technique used in
Ref.~\cite{bauberger}, which is makes use of the basic dispersion relation for the
one-loop self-energy function $B_0$, 
\begin{align}
B_0(p^2,m_1^2,m_2^2) &= \int_{(m_1+m_2)^2}^\infty {\rm d}\sigma \,
  \frac{\Delta B_0(\sigma,m_1^2,m_2^2)}{\sigma - p^2-i\epsilon}, \label{b0disp} \\
\Delta B_0(\sigma,m_1^2,m_2^2) &\equiv \frac{1}{\pi} \text{Im} \,
 B_0(\sigma,m_1^2,m_2^2) = (4\pi\mu^2)^{4-D} \,
  \frac{\Gamma(D/2-1)}{\Gamma(D-2)} \, \frac{\lambda^{(D-3)/2}(\sigma,m_1^2,m_2^2)}%
  {\sigma^{D/2-1}},
\end{align}
where $D$ is the space-time dimension and $\lambda(a,b,c) = (a-b-c)^2 - 4bc$.

\begin{figure}[tb]
\centering
\epsfig{figure=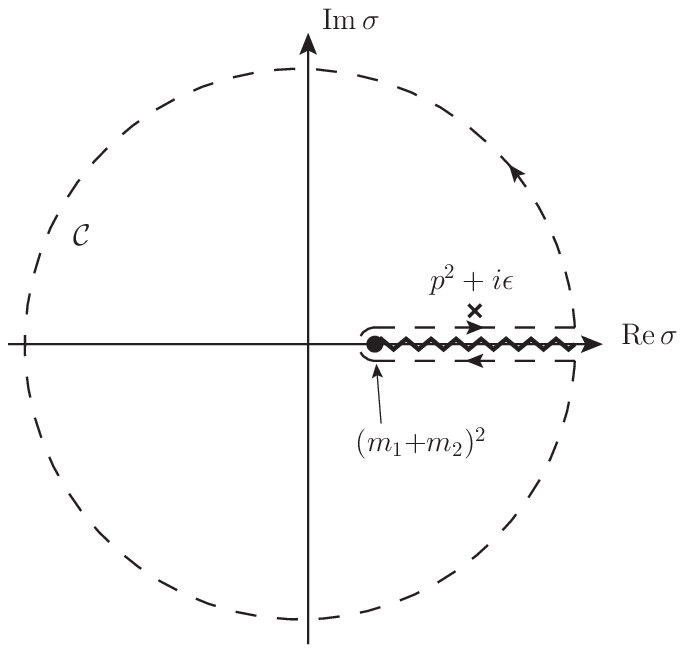, width=7cm, bb=130 400 456 706}
\hspace{1cm}
\epsfig{figure=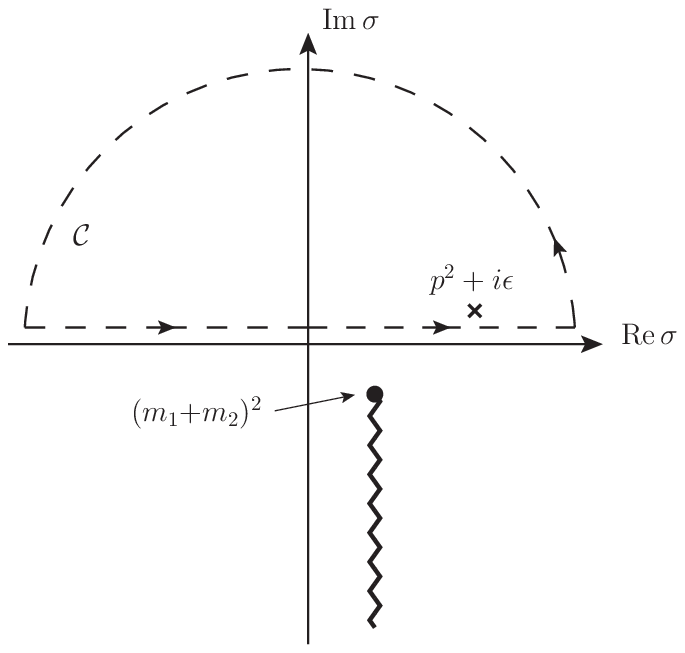, width=7cm, bb=130 400 456 706}
\mycaption{Integration contours for the dispersion relations for the one-loop
scalar self-energy function $B_0$ for the cases $m_1^2,m_2^2>0$ (left) and
$m_1^2<0$, $m_2^2>0$ (right). The zigzag lines denote the branch cuts, ending at
the branch point $(m_1{+}m_2)^2$. The circle sections are understood to have a radius $R \to \infty$.}
\label{fig:disp1}
\end{figure}

This dispersion relation is derived from the analytical properties of the $B_0$
function: For complex $p^2$, $B_0(p^2,m_1^2,m_2^2)$ has a branch point at
$p^2=(m_1+m_2)^2$, with the branch cut on the real-axis interval
$((m_1+m_2)^2,\infty)$. When using Cauchy's theorem, $B_0(p^2,m_1^2,m_2^2) =
\frac{1}{2\pi i} \oint_{\cal C} d\sigma \, \frac{B_0(\sigma,m_1^2,m_2^2)}{\sigma -
p^2-i\epsilon}$, one must choose a contour $\cal C$ that circumvents the branch
cut, as illustrated in Fig.~\ref{fig:disp1}~(left).
The discontinuity $\Delta B_0$ accounts for the difference of
$B_0(\sigma,m_1^2,m_2^2)$ for values of $\sigma$ just below and just above the
branch cut. The contour is closed with a circle at infinity, which gives
vanishing contribution for sufficiently small dimension $D$.

%------------------------------------------------------------------------------
%------------------------------------------------------------------------------

In order to apply this relation to the planar box diagram, it is
useful to introduce Feynman parameters for the propagators that depend only on
loop momemtum $q_2$ \cite{Awramik:2006uz}:
\begin{align}
&\frac{1}{[q_2^2-\mt^2][(q_2+k_1)^2-\mt^2][(q_2+k_1+k_2)^2-\mt^2]} = \int_0^1 dx \int_0^{1-x} dy \; \frac{2}{[(q_2+k')^2-m'^2]^3}, 
\\[2ex]
&k' = (1-x)k_1 + y\,k_2, \qquad
m'^2 = \mt^2 - xy(k_1+k_2)^2 - (1-x-y)(x k_1^2 + y k_2^2).
\end{align}
Then the $q_2$ loop can be expressed as
\begin{align}
&\int dx\,dy \int d^Dq_2 \; \frac{2}{[(q_1-q_2)^2-m_{q'}^2][(q_2+k')^2-m'^2]^3}
 \notag \\
&\quad = \int dx\,dy \; \frac{\partial^2}{\partial (m'^2)^2}
  \int_{\sigma_0}^\infty d\sigma \, 
  \frac{\Delta B_0(\sigma,m'^2,m_{q'}^2)}{\sigma - \tilde{q}_1^2} \notag \\
&\quad = \int dx\,dy \; \biggl\{
   \int_{\sigma_0}^\infty d\sigma \, 
   \frac{\partial^2_{m'}\Delta B_0(\sigma,m'^2,m_{q'}^2)}{\sigma - \tilde{q}_1^2}
	-\dfrac{m'+m_{q'}}{m'} \biggl[\frac{\partial_{m'}\Delta B_0(\sigma,m'^2,m_{q'}^2)}{\sigma - \tilde{q}_1^2}
   \biggr]_{\sigma \to \sigma_0} \biggr\}, \label{disp2a}
\intertext{where we have introduced the short-hand notation}
&\sigma_0 = (m'+m_{q'})^2, \qquad
\tilde{q}_1 = q_1+k'+i\epsilon, \qquad
\partial_{m'} = \frac{\partial}{\partial (m'^2)}\,,
\end{align}
and used the fact that $\Delta B_0(\sigma_0, m'^2,m_{q'}^2) = 0$. Unfortunately,
the $\sigma$ integral blows up at the lower boundary, and the term in [~] is also
divergent for $\sigma \to \sigma_0$, 
whereas only the sum of the two is finite.
To circumvent this problem, one can modify the integrand according to
\begin{align}
\int dx\,dy \; \biggl\{
 &\int_{\sigma_0}^\infty d\sigma \; \partial^2_{m'}\Delta
 B_0(\sigma,m'^2,m_{q'}^2)\,
 \biggl(\frac{1}{\sigma - \tilde{q}_1^2} - \frac{\sigma_0}{\sigma(\sigma_0 - \tilde{q}_1^2)}
 \biggr) \notag \\
 &+ \frac{\sigma_0}{\sigma_0 - \tilde{q}_1^2} \, 
 \partial^2_{m'}B_0(0,m'^2,m_{q'}^2) \biggr\}. \label{disp2b}
\end{align}
Here the extra term in the integrand of $\int d\sigma$ is added back in
integrated form, where the function $\partial^2_{m'}B_0$ can be expressed in
terms of basic logarithms (see appendix). With the modified integrand, the
boundary term in eq.~\eqref{disp2a} evaluates to zero.

Inserting eq.~\eqref{disp2b} into the remainder of the $q_1$ loop integral, one
obtains
\begin{align}
I_{\rm plan} = -\!\int dx \, dy \; \biggl\{ 
 &\int_{\sigma_0}^\infty d\sigma \; \partial^2_{m'}\Delta B_0(\sigma,m'^2,m_{q'}^2)
  \notag \\[-1ex]
 &\hspace{6em}\times \Bigl[ D_0(p_1^2,p_2^2,k'^2_2,k'^2_1,s,t',m_{V_1}^2,m_{f'}^2,m_{V_2}^2,\sigma)
  \notag \\
 &\hspace{8em} - \frac{\sigma_0}{\sigma}\,
  D_0(p_1^2,p_2^2,k'^2_2,k'^2_1,s,t',m_{V_1}^2,m_{f'}^2,m_{V_2}^2,\sigma_0)
  \Bigr] \notag \\
 &+\sigma_0 \; \partial^2_{m'}B_0(0,m'^2,m_{q'}^2) \,
 D_0(p_1^2,p_2^2,k'^2_2,k'^2_1,s,t',m_{V_1}^2,m_{f'}^2,m_{V_2}^2,\sigma_0)
 \biggr\}, \label{planr}
\end{align}
where $s = (p_1+p_2)^2$, $t' = (p_1-k'_1)^2$, and $D_0$ is the well-known
scalar one-loop box function \cite{hv,ff,denner}.

Since the double box diagrams are UV finite, all expressions in
eq.~\eqref{planr} can be computed for $D{=}4$ dimensions.

\medskip

The full diagram respresented by Fig.~\ref{fig:diags} (top-left) contains
additional terms with momenta $q_{1,2}$ in the numerator stemming from the Dirac
propagators and vertex structures. For terms depending on $q_2$, it is
convenient to perform a Passarino-Veltman decomposition
\cite{Passarino:1978jh,denner} of
$\partial^2_{m'}B_0(\tilde{q}_1^2,m'^2,m_{q'}^2)$ \emph{after} introduction of the
Feynman parameters. As a first step, let us shift the integration momentum to
$q'_2 \equiv q_2 + k'$:
\begin{align}
&\int d^4q_2 \; \frac{q_2^\mu q_2^\nu \cdots}{[(q_2+k')^2-m'^2]^3[(q_1-q_2)^2-m_{q'}^2]}
= \int d^4q'_2 \; \frac{(q'_2-k')^\mu (q'_2-k')^\nu \cdots}
 {[q'^2_2-m'^2]^3[(q'_2-q_1-k')^2-m_{q'}^2]}\,.
\end{align}
The terms with powers of $q'_2$ in the numerator can be decomposed
according to
\begin{align}
&\int d^4q'_2 \; \frac{q'^\mu_2}{[q'^2_2-m'^2]^3[(q'_2-q_1-k')^2-m_{q'}^2]}
 = -\tilde{q}_1^\mu \; \partial^2_{m'}B_1(\tilde{q}_1^2,m'^2,m_{q'}^2)\,, \notag \\
&\int d^4q'_2 \; \frac{q'^\mu_2 q'^\nu_2}{[q'^2_2-m'^2]^3[(q'_2-q_1-k')^2-m_{q'}^2]}
 \label{q2tens} \\
 &\quad = g^{\mu\nu} \; \partial^2_{m'}B_{00}(\tilde{q}_1^2,m'^2,m_{q'}^2) +
 \tilde{q}_1^\mu\tilde{q}_1^\nu \; \partial^2_{m'}B_{11}(\tilde{q}_1^2,m'^2,m_{q'}^2)\,,
 \notag \\
&\text{etc.} \notag
\end{align}
Each of the Passarino-Veltman functions
$\partial^2_{m'}B_{ij...}((q_1+k')^2,m'^2,m_{q'}^2)$ can then be represented
through a dispersion relation in the same manner as above:
\begin{align}
\partial^2_{m'}B_{ij...}(\tilde{q}_1^2,m'^2,m_{q'}^2) &= \frac{1}{\pi}
 \int_{\sigma_0}^\infty {\rm d}\sigma \;
  [\text{Im} \, \partial^2_{m'}B_{ij...}(\sigma,m'^2,m_{q'}^2)]\,
  \biggl[\frac{1}{\sigma - \tilde{q}_1^2} - \frac{\sigma_0}{\sigma(\sigma_0 - \tilde{q}_1^2)}
 \biggr] \notag \\
 &\quad + \frac{\sigma_0}{\sigma_0 - \tilde{q}_1^2} \, 
 \partial^2_{m'}B_{ij...}(0,m'^2,m_{q'}^2)\,.
\end{align}
Explicit expressions for $\text{Im} \,
\partial^2_{m'}B_{ij...}(\sigma,m_1^2,m_2^2)$ and
$\partial^2_{m'}B_{ij...}(0,m'^2,m_{q'}^2)$ are collected in the appendix.

Similarly, the $q_1$ loop will in general contain terms with different powers of
$q_1$ in the numerator, some of which in fact originate from eq.~\eqref{q2tens}.
These lead to Passarino-Veltman functions $D_1$, $D_2$, $D_3$, $D_{00}$, etc.
\cite{Passarino:1978jh},
which can be evaluated numerically by using, for example, the techniques
introduced in Ref.~\cite{ff,vanOldenborgh:1990yc}. In some cases, there are cancellations between
terms in the numerator and denominator, resulting in
$C_0,\,C_1,\,C_2,\,C_{00},\,...$ and $B_0,\,B_1,\,B_{00},\,...$ functions.

%%%%%%%%%%%%%%%%%%%%%%%%%%%%%%%%%%%%%%%%%%%%%%%%%%%%%%%%%%%%%%

\subsection{Non-planar diagrams}

The approach in the previous sub-section can be adapted also to the case of
non-planar box diagrams. Let us begin with the scalar non-planar integral
corresponding to the diagram in Fig.~\ref{fig:diags}~(top-right):
\begin{align}
I_{\rm np} = \int &d^Dq_1 \, d^Dq_2 \; \frac{1}{[q_1^2-m_{V_1}^2]
 [(q_1+p_1)^2-m_{f'}^2][(q_1+p_1+p_2)^2-m_{V_2}^2]} \notag \\
&\times \frac{1}{[(q_1-q_2)^2-m_{q'}^2][(q_1-q_2+k_1)^2-m_{q'}^2]
 [q_2^2-\mt^2][(q_2+k_2)^2-\mt^2]}\,.
\end{align}
Introducting two Feynman parameters, the $q_2$ loop can be written as
\begin{align}
&\int d^Dq_2 \; \frac{1}{[(q_1-q_2)^2-m_{q'}^2][(q_1-q_2+k_1)^2-m_{q'}^2]
 [q_2^2-\mt^2][(q_2+k_2)^2-\mt^2]} \notag \\
&\quad = \int_0^1 dx \int_0^1 dy \int d^Dq_2 \; 
 \frac{1}{[(q_1-q_2+(1-x)k_1)^2-m'^2_1]^2[(q_2+yk_2)^2-m'^2_2]^2}
 \notag \\
&\quad = \int dx\,dy \; \partial_{m'_1} \partial_{m'_2}
  \int_{\sigma_0}^\infty d\sigma \, 
  \frac{\Delta B_0(\sigma,m'^2_1,m'^2_2)}{\sigma - \tilde{q}_1^2} \notag \\
&\quad \begin{aligned}[b]  = \int dx\,dy \; \biggl\{
 &\frac{1}{\pi} \int_{\sigma_0}^\infty d\sigma \; 
 [\text{Im} \,\partial_{m'_1} \partial_{m'_2} B_0(\sigma,m'^2_1,m'^2_2)]\,
 \biggl(\frac{1}{\sigma - \tilde{q}_1^2} - \frac{\sigma_0}{\sigma(\sigma_0 - \tilde{q}_1^2)}
 \biggr) \\
 &+ \frac{\sigma_0}{\sigma_0 - \tilde{q}_1^2} \, 
 \partial_{m'_1} \partial_{m'_2} B_0(0,m'^2_1,m'^2_2) \biggr\},
\end{aligned} \label{dispnp}
\intertext{where}
&m'^2_1 = m_{q'}^2 - x(1-x)k_1^2, \qquad
m'^2_2 = \mt^2 - y(1-y)k_2^2, \qquad
\sigma_0 = (m_1'+m'_2)^2, \notag \\
&\tilde{q}_1 = q_1 + (1-x)k_1 + yk_2 + i\epsilon.
\end{align}
In the last step of eq.~\eqref{dispnp}, a threshold subtraction has again been 
utilized to ensure that the $\sigma$ integral is convergent at the lower
boundary.
Together with the $q_1$ integral, $I_{\rm np}$ then becomes
\begin{align}
I_{\rm np} = -\!\int &dx \, dy \; \biggl\{ 
 \frac{1}{\pi} \int_{\sigma_0}^\infty d\sigma \; 
 [\text{Im} \,\partial_{m'_1} \partial_{m'_2} B_0(\sigma,m'^2_1,m'^2_2)]
  \notag \\[-1ex]
 &\hspace{8em}\times \Bigl[ D_0(p_1^2,p_2^2,k'^2_2,k'^2_1,s,t',m_{V_1}^2,m_{f'}^2,m_{V_2}^2,\sigma)
  \notag \\
 &\hspace{10em} -\frac{\sigma_0}{\sigma}
  D_0(p_1^2,p_2^2,k'^2_2,k'^2_1,s,t',m_{V_1}^2,m_{f'}^2,m_{V_2}^2,\sigma_0)
  \Bigr]\notag \\
 &+\sigma_0 \; \partial_{m'_1} \partial_{m'_2} B_0(0,m'^2_1,m'^2_2) \,
 D_0(p_1^2,p_2^2,k'^2_2,k'^2_1,s,t',m_{V_1}^2,m_{f'}^2,m_{V_2}^2,\sigma_0)
 \biggr\}, \label{npr}
\end{align}
where all components of the integrands are well-known analytical functions.

\medskip

As before, the extension to tensor integrals can be realized by using dispersion
relations for $\partial_{m'_1} \partial_{m'_2} B_1$, $\partial_{m'_1}
\partial_{m'_2} B_{00}$, etc. (see appendix for explicit formulas). Similarly,
the usual Passario-Veltman tensor functions $D_1$, $D_2$, etc. can be used  for
tensor structures in the $q_1$ loop integrals.

\medskip

An additional complication arises for the non-planar diagram with $W$ bosons, in
which case $m_{q'} = \mb \approx 0$, so that $m'^2_1$ is negative. As a result,
the branch point $\sigma_0$ of the $\partial_{m'_1} \partial_{m'_2} B_0$
function is in the lower complex half-plane rather than on the real axis (see
Fig.~\ref{fig:disp1}), so that the
dispersion relation \eqref{b0disp} must be modified. One option is to choose a
contour along the real axis, which is closed via a semi-circle in the upper
complex half-plane, leading to
\begin{align}
\partial_{m'_1} \partial_{m'_2} B_0(p^2,m'^2_1,m'^2_2) &= 
  \frac{1}{2\pi i} \int_{-\infty}^\infty {\rm d}\sigma \,
  \frac{\partial_{m'_1} \partial_{m'_2} B_0(\sigma,m'^2_1,m'^2_2)}
  {\sigma - p^2-i\epsilon}.
\end{align}
Using this relation, $I_{\rm np}$ can be expressed as
\begin{align}
I_{\rm np} = -\!\int &dx \, dy \; 
 \int_{-\infty}^\infty d\sigma \; 
 \frac{\partial_{m'_1} \partial_{m'_2} B_0(\sigma,m'^2_1,m'^2_2)}{2\pi i}
  \,D_0(p_1^2,p_2^2,k'^2_2,k'^2_1,s,t',m_{V_1}^2,m_{f'}^2,m_{V_2}^2,\sigma-i\epsilon)
  \label{inp} .
\end{align}
The $i\epsilon$ in the last mass parameter of the $D_0$ function is important to
properly define its result for all values of $\sigma$. In fact, as also
discussed in the next section, it turns out to be necessary to include a small
numerical value for $\epsilon$ when using {\sc LoopTools} \cite{looptools} for
the evaluation of certain Passarino-Veltman functions.

%%%%%%%%%%%%%%%%%%%%%%%%%%%%%%%%%%%%%%%%%%%%%%%%%%%%%%%%%%%%%%

\section{Implementation and numerical results}
\label{res}

In this section we describe how the approach described in the previous section
has been applied to the calculation of all box diagrams of the form in
Fig.~\ref{fig:diags}. The results presented in this section are based on two
independent realizations of the calculation, in order to enable cross-checks
between the two.

Both implementations employ {\sc Mathematica} \cite{mathematica} as the
framework for algebraic manipulations and {\sc FeynArts 3} \cite{feynarts} for
the generation of diagrams and amplitudes in Feynman gauge. One implementation uses {\sc FeynCalc
9} \cite{feyncalc} for carrying out the Lorentz and Dirac algebra and then
divides the expressions into individual tensor integral terms, as discussed at
the end of section~\ref{plan}. Each of these terms is integrated separately
within C++, using the {\sc LoopTools 2.15} \cite{looptools} package for the
Passarino-Veltman functions and the adaptive Gaussian quadrature integration
routine from the {\sc Boost} library \cite{boost}. The integration results are
then added up to obtain full diagram results.

The second implementation performs the Lorentz and Dirac algebra with in-house
routines and then tranforms the expressions for complete diagrams into a single
integrand each. The numerical integration is carried out in C++ using the
adaptive Gauss-Kronrod integration routine from TVID \cite{tvid}, which is based
on the {\sc Quadpack} library \cite{quadpack}. It also uses {\sc LoopTools} for
the Passarino-Veltman functions in the integrand.

In light of the fact that the double-box integrals are UV-finite, it is
advantageous to perform the Lorentz and Dirac algebra in 4 dimensions, thus
avoiding any ambiguities in the treatment of $\gamma_5$. Even though the sum of
all box diagrams considered here is IR finite, individual diagrams with photons
are IR divergent, and thus an IR regulator is required. A convenient choice is
the use of a small photon mass, $m_\gamma$, since it is trivially compatible
with the 4-dimensional Lorentz and Dirac algebra.

It is advantageous to implement the three-dimensional numerical integrals in a
nested structure, with the $\sigma$ integral being the inner-most integral,
since this makes the adaptive integration algorithms most effective. The
achievable precision is limited by the double precision floating point algebra
used in the default compilation of {\sc LoopTools}. In fact, numerical
instabilities are typically encountered near the lower and upper limits of the
$\sigma$ integration. These can be mitigated by introducing cut-offs at both
ends,
\begin{align}
\int_{\sigma_0}^\infty d\sigma \quad\to\quad
\int_{\sigma_0(1+\delta)}^\Lambda d\sigma ,
\end{align}
where $\delta \ll 1$ and $\Lambda$ should be much larger than all mass and
momentum scales in the matrix element. The error due to these cut-offs can be
further mitigated by observing that the integrand approximately
behaves like $\sim (\sigma-\sigma_0)^{-1/2}$ near the lower threshold and
$\sim (A+B\log \sigma)/\sigma^2$ for large $\sigma$. Thus one can introduce
additional correction terms,
\begin{align}
\int_{\sigma_0}^\infty d\sigma \; f(\sigma) \quad\to\quad
\int_{\sigma_0(1+\delta)}^\Lambda d\sigma \; f(\sigma)
 + 2\sigma_0\delta\,f(\sigma_0\delta)+ \Lambda \, f(\Lambda).
\end{align}
For the two-loop box diagrams considered
here, suitable choices for $\delta$ and $\Lambda$ are ${\cal
O}(10^{-4}...10^{-3})$ and ${\cal O}(10^8...10^{12}\gev^2)$, respectively.
One can verify that the integration result does not change very much when 
varying $\delta$ and $\Lambda$ within one order of magnitude, and this variation
can be interpreted as a source of uncertainty for the final results (see below).

For the non-planar diagrams, additional instabilities occur for $x \approx y$,
when the Gram determinants for some Passarino-Veltman tensor functions vanish.
Our two implementations use two different strategies for mitigating this
problem: (a) splitting one of the two integration intervals, such that none
of the Gaussian points of the $x$ integration lies too close to the ones for the
$y$ integration; or (b) interpolating the $y$ integration across a small
interval, $y \in [x-\Delta x, x+\Delta x]$. A reasonable comprise between
accuracy and stability is achieved for $\Delta x \sim {\cal O}(10^{-2})$. Both
methods yield consistent results, and the impact of varying $\Delta x$ by a
factor of a few can be used as a contribution to the final error estimate.

Finally, the evaluation of the non-planar $WW$ box requires 
an explicit value for the Feynman $i\epsilon$, see eq.~\eqref{inp},
to avoid instabilities in {\sc LoopTools} for negative $\sigma$. A value of
$\epsilon = 10^{-9}|\sigma|$ is chosen for the results presented
below.

\medskip

In the following, numerical results will be presented for the different classes
of box diagrams, which are distinguished by the gauge bosons $V_{1,2}$ appearing
inside the loops. The numbers are obtained by contracting the matrix elements
for the two-loop box diagrams, ${\cal M}_2$, with the tree-level matrix element
${\cal M}_0$, averaging over $e^\pm$ helicities and summing over the final-state
$Z$-boson polarization states.

\begin{figure}[tb]
\vspace{1ex}
\centering
\begin{tabular}{ll}
\epsfig{figure=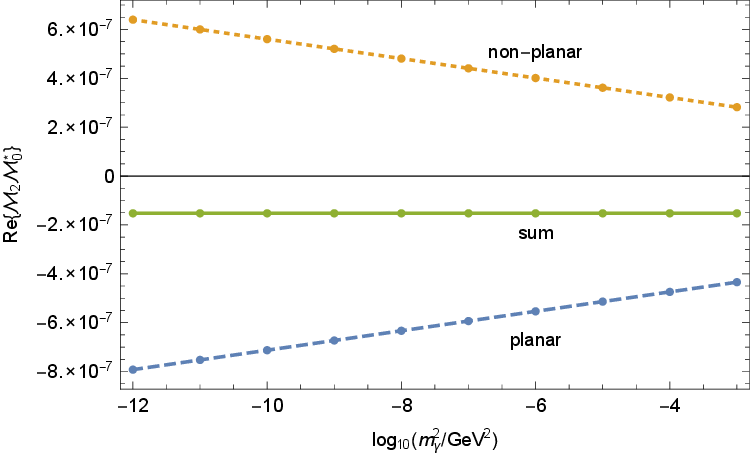, height=1.85in} 
 &
\epsfig{figure=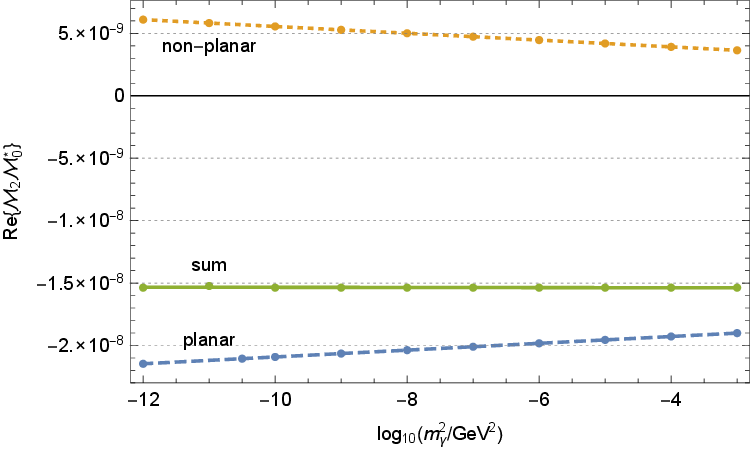, height=1.85in}
\end{tabular}
\mycaption{Dependence of the $\gamma\gamma$ (left)
and $\gamma Z$ (right) two-loop boxes on the photon mass $m_\gamma$. The lines 
depict linear fits to the data points.}
\label{fig:mgamma}
\end{figure}

%-------------------------------------------------------------
\begin{table}[tb]
\vspace{1ex}
\centering
\renewcommand{\arraystretch}{1.2}
\begin{tabular}{c|c|c|}
\cline{2-3}
(a) & Parameter \ & Value \\
\cline{2-3}
& \MZ & 91.1876~GeV \\
& \MW & 80.379~GeV \\
& \MH & 125.1~GeV \\
& \mt & 172.76~GeV \\
& $\alpha$ & \phantom{1}1/137 \\
& $E_{\rm CM}$ & 240~GeV \\
\cline{2-3}
\end{tabular}
\hspace{.5in}
\begin{tabular}{c|c|l|}
\cline{2-3}
(b) & $V_1V_2$ diagr.\ class \ & \multicolumn{1}{c|}{Re$\{{\cal M}_2{\cal M}_0^*\}$} \\
\cline{2-3}
& $\gamma\gamma$ & $-1.524(1) \times 10^{-7}$ \\
& $\gamma Z$ & $-1.537(1) \times 10^{-8}$ \\
& $ZZ$ planar & $-4.402(4)\times 10^{-8}$ \\
& $ZZ$ non-planar & $\phantom{-}1.724(2)\times 10^{-8}$ \\
& $WW$ planar & $-1.1392(8)\times 10^{-6}$ \\
& $WW$ non-planar & $-5.577(5)\times 10^{-7}$ \\
\cline{2-3}
\end{tabular}
\mycaption{(a) Input values used for the numerical examples. (b) Results for
different classes of two-loop box diagrams, distinguished by topology and internal gauge-boson species,
for scattering angle \mbox{$\theta = \pi/2$}. The numbers in brackets indicate
an estimate of the intrinsic uncertainty in the last shown digit (see text for
more details).}
\label{tab:data1}
\end{table}
%-------------------------------------------------------------
Using the inputs in Tab.~\ref{tab:data1}~(a), we obtain the numbers shown in
Tab.~\ref{tab:data1}~(b). For the diagrams with photons, the dependence on the photon
mass regulator only drops out when adding planar and non-planar diagrams, as
illustrated in Fig.~\ref{fig:mgamma}. Also shown in Tab.~\ref{tab:data1}~(b) is
an esimate of the precision, as obtained by varying the lower and upper cut-off
of the $\sigma$ by one order of magnitude each. For the non-planar diagrams,
the impact of varying the width $\Delta x$ of the window around $y=x$ by a
factor 2 is also considered. The integration times for each line in
Tab.~\ref{tab:data1}~(b) range from a few minutes up to about half an hour on a
single CPU core.
%-------------------------------------------------------------
%-------------------------------------------------------------

\begin{figure}[tb]
\vspace{1ex}
\centering
\epsfig{figure=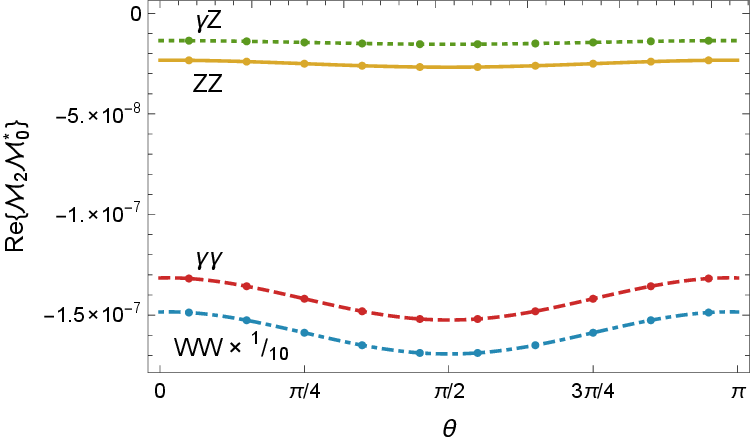, height=2.5in} 
\vspace{-1ex}
\mycaption{Dependence of various groups of two-loop box contributions on the
scattering angle $\theta$.}
\label{fig:diff}
\end{figure}

Figure~\ref{fig:diff} shows the dependence on the scattering angle $\theta$. 
The differential distributions are symmetric, since each subset of box diagrams
has a $t\leftrightarrow u$ crossing symmetry, where $t,u$ are the usual Mandelstam
variables.
%-------------------------------------------------------------
%-------------------------------------------------------------

One can see that the diagrams with $W$ bosons produce results that are about one
order of magnitude larger than the ones with neutral bosons. This may be
explained by the fact than the effective $WWZH$ interaction (corresponding to
the fermion loop in our two-loop diagrams) can be represented by a dimension-6
operator, whereas $\gamma\gamma ZH$, $\gamma ZZH$ and $ZZZH$ interactions are
related to dimension-8 operators.
The diagrams with $Z$ bosons are additionally suppressed by the small $Zee$
vector coupling in the SM.

%%%%%%%%%%%%%%%%%%%%%%%%%%%%%%%%%%%%%%%%%%%%%%%%%%%%%%%%%%%%%%

\section{Summary}
\label{sum}

Box diagrams in $2\to 2$ process can be efficiently evaluated with a numerical
method that combines Feynman parametrization and a dispersion relation for one
sub-loop, while standard analytical expressions are used for the other sub-loop.
Tensor structures in the numerator can be handled by adjusting the dispersion
relation for the first loop
and using Passarino-Veltman reduction for the second sub-loop. The resulting
three-dimensional numerical integrals can be efficiently evaluated using
nested adaptive one-dimensional integration algorithms.

The efficacy of the technique has been demonstrated by computing planar and
non-planar box diagrams with top quarks contributing to the two-loop electroweak
corrections for the process $e^+e^- \to HZ$. Infrared divergencies from QED can
be controlled with a photon mass, without loss of numerical precision. Results
with a relative uncertainty of about 0.1\% can be obtained in a few minutes on a
single CPU core. The longest run-time (about half an hour) is required for
diagrams with a physical cut in the fermion sub-loop, which occurs for the
non-planar topology in the top-right of Fig.~\ref{fig:diags} with $V_{1,2}=W$
and $q'=b$. In this case a modified version of the dispersion relation is used,
with an integration contour along the entire real axis instead of just the
positive real axis. 

It should be noted that our current implementation of the other diagrams is
limited to center-of-mass energies below the $t\bar{t}$ threshold, $\sqrt{s} <
2\mt$, because otherwise a physical threshold would open up in the fermion
sub-loop there as well. Nevertheless, an extension to higher center-of-mass
energies could be achieved by using the modified dispersion relation for all
diagrams, even though it may come at the cost of a slight loss of accuracy and
increased integration time.

The numerical precision is primarily limited by the accuracy of the evaluation
of the basic one-loop Passarino-Veltman functions $B_1$, $C_{1,2}$,
$C_{00,11,12,22}$, $D_{1,2,3}$, etc. In our current implementation, {\sc LoopTools}
\cite{looptools} with double-precision floating point arithmetic is used for
this purpose. Improvements could be made by using quadruple precision numbers or
by performing expansions in the regions where numerical instabilities are
encountered. However, a relative precision of 0.1\% for the evaluation of
two-loop box diagrams is already sufficient for a range of important
phenomenological applications, including the $e^+e^- \to HZ$ process at future
Higgs factories.

The techniques described in this paper could, in principle, also be applied for
the calculation of electroweak corrections to other $2\to 2$ process, such as
$e^+e^- \to W^+W^-$.

%%%%%%%%%%%%%%%%%%%%%%%%%%%%%%%%%%%%%%%%%%%%%%%%%%%%%%%%%%%%%%

\section*{Acknowledgments}

The authors would like to thank Keping Xie for useful feedback on the
manuscript.
This work has been supported in part by the National Science Foundation under
grant no.\ PHY-1820760.

%%%%%%%%%%%%%%%%%%%%%%%%%%%%%%%%%%%%%%%%%%%%%%%%%%%%%%%%%%%%%%

\appendix
\section{Appendix: Integration kernels}

In the following, explicit expressions for the dispersion integration kernels
for various tensor integrals are listed. As before, we use the notation
$\partial_m \equiv \partial/\partial (m^2)$, and we also make use of 
the abbreviation $\lambda = \sigma^2 + m_1^4 + m_2^4 -
2(\sigma m_1^2 + \sigma m_2^2 + m_1^2m_2^2)$.
\begin{align}
\text{Im} \,\partial^2_{m_1}B_0(\sigma,m_1^2,m_2^2) &=
 -\pi\, \frac{4m_2^2}{\lambda^{3/2}}\,, \displaybreak[0] \\
\text{Im} \,\partial^2_{m_1}B_1(\sigma,m_1^2,m_2^2) &=
% \pi\, \frac{(m_1^2-m_2^2-\sigma)(\lambda-2m_2^2\sigma)}{\sigma^2\lambda^{3/2}}\,, 
 \pi\, \frac{4m_2^2\sigma^2 - (m_1^2-m_2^2-\sigma)(\lambda-2m_2^2\sigma)}
 {\sigma^2\lambda^{3/2}}\,, 
 \displaybreak[0] \\
\text{Im} \,\partial^2_{m_1}B_{00}(\sigma,m_1^2,m_2^2) &=
% -\pi\, \frac{\lambda+2m_2^2\sigma}{2\sigma^2\lambda^{1/2}}\,, 
 -\pi\, \frac{\lambda+2m_1^2\sigma}{2\sigma^2\lambda^{1/2}}\,, 
 \displaybreak[0] \\
\text{Im} \,\partial^2_{m_1}B_{11}(\sigma,m_1^2,m_2^2) &=
% \pi\, \frac{3\lambda^2 - (\lambda-2m_2^2\sigma)^2}{\sigma^3\lambda^{3/2}}\,, 
 2\pi\, \frac{\lambda[\lambda+\sigma(m_1^2+m_2^2-\sigma)] - 2\sigma^2 m_1^2
 m_2^2}{\sigma^3\lambda^{3/2}}\,,
 \displaybreak[0] \\
\text{Im} \,\partial^2_{m_1}B_{001}(\sigma,m_1^2,m_2^2) &=
% -\pi\, \frac{(m_1^2-m_2^2-\sigma)(\lambda+m_2^2\sigma)}{2\sigma^3\lambda^{1/2}}\,, 
 \pi\, \frac{(m_1^2-m_2^2)(\lambda+m_2^2\sigma)+m_2^2\sigma^2}{2\sigma^3\lambda^{1/2}}\,, 
 \displaybreak[0] \\
\text{Im} \,\partial^2_{m_1}B_{111}(\sigma,m_1^2,m_2^2) &=
 \frac{\pi}{\sigma^4\lambda^{3/2}}\, \bigl\{(m_2^2-m_1^2-\sigma)[3\lambda^2 + 4m_2^2\sigma\lambda
 \notag \\
 &\qquad +\sigma^2(3(m_1^2-m_2^2-\sigma)(m_1^2+m_2^2+\sigma)-2m_1^2m_2^2)]
 \notag \\
 &\qquad+12\sigma^3[m_1^4+m_2^4-\sigma(m_1^2+m_2^2)]\bigr\}\,, 
 \displaybreak[0] \\[1em]
\text{Im} \,\partial_{m_1}\partial_{m_2}B_0(\sigma,m_1^2,m_2^2) &=
 \pi\, \frac{2(m_1^2+m_2^2-\sigma)}{\lambda^{3/2}}\,, \displaybreak[0] \\
\text{Im} \,\partial_{m_1}\partial_{m_2}B_1(\sigma,m_1^2,m_2^2) &=
 \pi\, \frac{4m_1^2\sigma^2 - (m_2^2-m_1^2-\sigma)(\lambda-2m_1^2\sigma)}
 {\sigma^2\lambda^{3/2}}\,, 
 \displaybreak[0] \\
\text{Im} \,\partial_{m_1}\partial_{m_2}B_{00}(\sigma,m_1^2,m_2^2) &=
 \pi\, \frac{(m_1^2-m_2^2)^2-\sigma(m_1^2+m_2^2)}{2\sigma^2\lambda^{1/2}}\,, \displaybreak[0] \\
\text{Im} \,\partial_{m_1}\partial_{m_2}B_{11}(\sigma,m_1^2,m_2^2) &=
 \pi\, \frac{2m_1^2\sigma^2(m_1^2+m_2^2-\sigma)
  -\lambda[2\lambda + \sigma(3m_1^2+m_2^2-\sigma)]}{\sigma^3\lambda^{3/2}}\,.
\end{align}
The integrated functions for zero momentum are given by, in terms of $r =
m_2^2/m_1^2$,
\begin{align}
\partial^2_{m_1}B_0(0,m_1^2,m_2^2) &= \frac{m_1^{-4}}{(1-r)^2}
 \biggl[1+r+2r\frac{\ln r}{1-r}\biggr], \displaybreak[0] \\
\partial^2_{m_1}B_1(0,m_1^2,m_2^2) &= \frac{m_1^{-4}}{2(1-r)^3}
 \biggl[-1-5r-2r(2+r)\frac{\ln r}{1-r}\biggr], \displaybreak[0] \\
\partial^2_{m_1}B_{00}(0,m_1^2,m_2^2) &= \frac{m_1^{-2}}{4(1-r)^2}
 \biggl[-1+3r+2r^2\frac{\ln r}{1-r}\biggr], \displaybreak[0] \\
\partial^2_{m_1}B_{11}(0,m_1^2,m_2^2) &= \frac{m_1^{-4}}{3(1-r)^4}
 \biggl[1+10r+r^2+6r(1+r)\frac{\ln r}{1-r}\biggr], \displaybreak[0] \\
\partial^2_{m_1}B_{001}(0,m_1^2,m_2^2) &= \frac{m_1^{-2}}{12(1-r)^3}
 \biggl[1-5r-2r^2-6r^2\frac{\ln r}{1-r}\biggr], \displaybreak[0] \\
\partial^2_{m_1}B_{111}(0,m_1^2,m_2^2) &= \frac{m_1^{-4}}{12(1-r)^5}
 \biggl[-3-47r-11r^2+r^3-12r(2+3r)\frac{\ln r}{1-r}\biggr], 
 \displaybreak[0] \\[1em]
\partial_{m_1}\partial_{m_2}B_0(0,m_1^2,m_2^2) &= \frac{m_1^{-4}}{(1-r)^2}
 \biggl[-2-(1+r)\frac{\ln r}{1-r}\biggr], \displaybreak[0] \\
\partial_{m_1}\partial_{m_2}B_1(0,m_1^2,m_2^2)) &= \frac{m_1^{-4}}{2(1-r)^3}
 \biggl[5+r+(2+4r)\frac{\ln r}{1-r}\biggr], \displaybreak[0] \\
\partial_{m_1}\partial_{m_2}B_{00}(0,m_1^2,m_2^2) &= \frac{m_1^{-2}}{4(1-r)^2}
 \biggl[-1-r-2r\frac{\ln r}{1-r}\biggr], \displaybreak[0] \\
\partial_{m_1}\partial_{m_2}B_{11}(0,m_1^2,m_2^2) &= \frac{m_1^{-4}}{6(1-r)^4}
 \biggl[-17-8r+r^2-6(1+3r)\frac{\ln r}{1-r}\biggr].
\end{align}

%%%%%%%%%%%%%%%%%%%%%%%%%%%%%%%%%%%%%%%%%%%%%%%%%%%%%%%%%%%%%%

\end{document}